\documentclass[amsmath,superscriptaddress,citeautoscript,twocolumn,
showpacs,floatfix,prb]{revtex4}
\usepackage{amssymb,amsmath}
\usepackage{graphics}
\usepackage{epstopdf}
\usepackage{epsfig}
\usepackage{color}
\usepackage{amsfonts}
\usepackage{bm}
\usepackage{amscd}

\begin{document}

\title{Quantum coherence of the molecular states and their corresponding currents in nanoscale Aharonov-Bohm interferometers}

\author{Jian-Heng Liu}
\affiliation{Department of Physics and Centre for Quantum Information Science, National Cheng Kung University, Tainan 70101, Taiwan}

\author{Matisse Wei-Yuan Tu}
\affiliation{Department of Physics and Centre for Quantum Information Science, National Cheng Kung University, Tainan 70101, Taiwan}
\affiliation{Department of Physics and Center of Theoretical and Computational Physics, University of Hong Kong, Hong Kong}

\author{Wei-Min Zhang}
\email{wzhang@mail.ncku.edu.tw}
\affiliation{Department of Physics and Centre for Quantum Information Science, National Cheng Kung University, Tainan 70101, Taiwan}

\begin{abstract}
By considering a nanoscale Aharonov-Bohm (AB) interferometer containing a parrallel-coupled double dot coupled to the source and drain electrodes,
we investigate the AB phase oscillations of transport current via the bonding and antibonding state channels. The results we obtained
justify the experimental analysis given in [Phys. Rev. Lett. \textbf{106}, 076801 (2011)] that bonding state currents in different energy configurations are almost the same.
On the other hand, we extend the analysis to the transient transport current components flowing through different channels, to explore
the effect of the parity of bonding and antibonding states on the AB phase dependence of the corresponding current components in the transient regime.
The relations of the AB phase dependence between the quantum states and the associated current components are analyzed in details,
which provides useful information for the reconstruction of quantum states through the measurement of the transport current in such systems.
With the coherent properties in the quantum dot states as well as in the transport currents, we also provide a way to manipulate the bonding and antibonding states by the AB magnetic flux.

\end{abstract}
\maketitle

\section{Introduction}

Quantum coherence of electrons in nanostructures is expected to manage quantum computation and quantum information.
It is essential to prepare and read out the state of the qubit in quantum information processing.
There have been many experiments and theoretical analyses on quantum coherence manipulation of electron states in DQDs which are thought to be a promising charge qubit.\cite{PCQ1998,QSTnQD2003,QSTnQD20032,QSTnQD2005,QSTnQD20052,QSTnQD2006,QSTnQD2007,Fol2009,QSTnQD2010,Maune2012,QSTnQD2013,QSTnQD2014,PCQ32005,PCQ2004} 
The techniques to reconstruct quantum states from series of measurements about the system are known as quantum state tomography.\cite{QSTnQD20062,Med2013,QSTnQD20142}
Quantum state tomography is resource demanding and it aims at very detailed description of coherence of quantum states.
On the other hand, transport measurement utilizing quantum interference has revealed the main coherent properties of traveling electrons.
How the latter can be associated with the coherence of local quantum states of the DQDs is worthy further investigation.

Quantum coherence can be detected through the Aharonov-Bohm (AB) interference.\cite{ABeffect1959}
Double quantum dots (DQDs) embedded in AB geometry were achieved in Ref.~[\onlinecite{ABwithDQDs2001,ABwithDQDs2004,ABwithDQDs22004}].
The results show that the phase coherence is also maintained in these devices.
The phase coherence of electrons through each dot would induce oscillating current as a function of the magnetic flux, which is simply called the AB oscillation in the literature. In Coulomb blockade and cotunneling regimes, it is predicted theoretically that currents through spin-singlet and triplet states carry AB phases with half period difference. \cite{DP2000}
For one-electron states, the AB phase difference by half a period is also anticipated in currents through bonding state (BS) and antibonding state (AS) channels,\cite{DP2004,DP2006} which has been detected experimentally.\cite{DetectParity2011}
This AB phase difference is thought to be resulted from the parity of the AS and BS wave functions, which is a property of the device geometry.
In Ref.~[\onlinecite{DetectParity2011}], the authors determine the parities of AS and BS by measuring the electron currents flowing through the corresponding channels.
Inspired by the ability of detecting currents from different channels, we are able to investigate the coherent properties of quantum states in DQDs, for which the corresponding currents should have a direct connection.
In Ref.~[\onlinecite{DetectParity2011}], two different energy configurations are used, which are succeeded by two different gate voltage settings.
The transport currents under these two configurations are measured.
The measured currents are used to determine the AS and/or BS channels in one of the configurations.
We would verify the validity of this analysis using our theoretical framework of the quantum transport theory based on master equation approach.\cite{CGFSF2008,TC2010,TC2012}
On the other hand, nanoscale AB interference has been discussed mostly in the steady-state regime.
Our previous works on transient quantum transport\cite{TC2010,TC2012,Yang14} can also be applied to discuss the formation of AS and BS in the transient regime.
We show the correspondence of AB phase dependences of the occupation probabilities of AS and BS and the transport currents of the corresponding channels, which should provide useful information for the reconstruction of the quantum states through the measurement of transport current.
Finally, we discuss the way to manipulate the AS and BS with the AB magnetic flux.

The rest of the paper is organized as follow:
In Sec.~\ref{s2}, we begin with the system of a DQDs coupled to two leads to study the time evolution of the reduced density matrix of the DQDs system and the transport currents flowing through AS and BS channels.
In Sec.~\ref{s3}, we obtain the condition for the validity of the method used in Ref.~[\onlinecite{DetectParity2011}] in analysing the connection of AS and BS with the measured currents.
We also extend such analysis to the transient transport regime.
In Sec.~\ref{s33}, we provide in details the correspondence of the AB phase dependence between the AS and BS density matrix elements with the associated transport currents.
A way to manipulate AS and BS density matrix elements with the AB magnetic flux is also given.
Finally, a summary is given in Sec.~\ref{s4}.

\section{Coupled DQDs molecule}
\label{s2}
We consider a nanosystem of two coupled single-level QDs coupled to two leads.
The Hamiltonian of this system is given by
\begin{equation}
H=H_{DQD}+H_{B}+H_{T},
\end{equation}
where $H_{DQD}$ is Hamiltonian of DQDs.
\begin{equation}
H_{DQD}=\sum_{i=1}^{2}\epsilon_{ij}d_{i}^{\dagger}d_{j},
\end{equation}
and $d_{i}\;(d_{i}^{\dagger})$ is annihilation (creation) operator
in $i$th QD, $\epsilon_{ii}$ is the energy level of $i$th QD and $\epsilon_{ij}$
with $i\neq j$ is the tunneling matrix element between the DQDs.
The Hamiltonian of the two leads is given by $H_{B}$:
\begin{equation}
H_{B}=\sum_{\alpha=L,R}\sum_{k}\varepsilon_{\alpha k}c_{\alpha k}^{\dagger}c_{\alpha k},
\end{equation}
where the label $\alpha$ denotes the left or right lead, and $c_{\alpha k}\;(c_{\alpha k}^{\dagger})$
is the annihilation (creation) operator of the $k$th level in lead $\alpha$.
The Hamiltonian $H_{T}$ describes the tunnelings between the QDs and the leads.
\begin{equation}
H_{T}=\sum_{\alpha=L,R}\sum_{i=1}^{2}\sum_{k}(V_{i\alpha k}d_{i}^{\dagger}c_{\alpha k}+h.c).
\end{equation}
By threading a magnetic flux $\Phi$ to the above system, the tunneling matrix elements would carry a AB phase, $V_{i\alpha k}=\bar{V}_{i\alpha k}e^{i\phi_{i\alpha}}$,
$\phi_{i\alpha}$ is the AB phase that electrons carry during the tunneling from $\alpha$ lead to $i$th dot, and $\bar{V}_{i\alpha k}$ is the
real tunneling amplitude.
The AB phase will also affect on $H_{DQD}$, i.e. for $i\neq j$, $\epsilon_{ij}=\bar{\epsilon}_{ij}e^{i\phi_{ij}}$
where $\bar{\epsilon}_{ij}=-t_{c}$ is a real amplitude and $\phi_{ij}$ is AB phase from $j$th dot to $i$th dot.
The relation of the AB phases with the magnetic flux  $\Phi$ is given by $\phi_{1L}-\phi_{1R}+\phi_{2R}-\phi_{2L}=2\pi \Phi/\Phi_{0}=\varphi$,
where $\Phi_0$ is the flux quanta.
We also set $\phi_{12}=0$ according to Ref.~[\onlinecite{DetectParity2011}].

\subsection{Exact master equation}
\label{matrix elements}
The system described above can be treated as an open quantum device.
The dynamics of the central system is described by the reduced density matrix $\rho (t)$, which is obtained by tracing over all the degrees of freedom of the reservoirs from the total density matrix $\rho_{tot}(t)$ of the total (the DQDs plus the leads), $\rho(t)=\mathrm{Tr}_{R}\left[ \rho_{tot}(t)\right]$.
The exact master equation which governs the dynamics of $\rho (t)$ for QDs has been derived:\cite{CGFSF2008}
\begin{align}
\label{MasterEq}
\frac{d\rho(t)}{dt} = & \frac{1}{i}\left[H_{S}^{\prime}(t),\rho(t)\right] +  \sum_{i,j} \Big\{\gamma_{ij}(t) \big[2d_{j}\rho(t)d_{i}^{\dagger} \nonumber \\
 & -d_{i}^{\dagger}d_{j}\rho(t)-\rho(t)d_{i}^{\dagger}d_{j}\big]  + \widetilde{\gamma}_{ij}(t)\big[d_{i}^{\dagger}\rho(t)d_{j} \nonumber \\
 & -d_{j}\rho(t)d_{i}^{\dagger}+d_{i}^{\dagger}d_{j}\rho(t)-\rho(t)d_{j}d_{i}^{\dagger}\big]\Big\},
\end{align}
where $H_{S}^{\prime}(t)=\sum_{i,j}\epsilon_{ij}^{\prime}(t)d_{i}^{\dagger}d_{j}$ is an effective Hamiltonian and $\epsilon_{ij}^{\prime}(t)$ is the renormalized time-dependent energy level ($i=j$) or the shifted interdot transition amplitude ($i\neq j$) between the DQDs.
All the time-dependent coefficients in Eq.(\ref{MasterEq}) are determined by the retarded Green function $\boldsymbol{u}(t,t_{0})$ and the correlation Green function $\boldsymbol{v}(\tau,t)$ in Keldysh's nonequilibrium Green function technique.\cite{TC2010}
Explicitly, the renormalized energy levels of DQDs $\boldsymbol{\epsilon}^{\prime}(t)$, the dissipation coefficient $\boldsymbol{\gamma}(t)$ and the fluctuation coefficient $\widetilde{\boldsymbol{\gamma}}(t)$ are given by:
\begin{subequations}
\begin{align}
& \boldsymbol{\epsilon}^{\prime}(t) =
  \frac{1}{2}\big[\dot{\boldsymbol{u}}(t)\boldsymbol{u}^{-1}(t)- {\rm H.c}\big],
\\
& \boldsymbol{\gamma}(t)  =  -\frac{1}{2}\big[\dot{\boldsymbol{u}}(t)\boldsymbol{u}^{-1}(t)+{\rm H.c}\big], \\
& \widetilde{\boldsymbol{\gamma}}(t)  = \dot{\boldsymbol{v}}(t) \!-\! \big[\dot{\boldsymbol{u}}(t)\boldsymbol{u}^{-1}(t)\boldsymbol{v}(t)+{\rm H.c}\big],
\end{align}
\end{subequations}
where $\boldsymbol{u}(t)\equiv \boldsymbol{u}(t,t_{0})$ and $\boldsymbol{v}(t)\equiv \boldsymbol{v}(t,t)$.
The Green function $\boldsymbol{u}(t,t_{0})$ obeys the following intergro-differential equation
\begin{equation}
\frac{\partial}{\partial t}\boldsymbol{u}(t,t_{0})+i\boldsymbol{\epsilon}\boldsymbol{u}(t,t_{0})+\sum_{\alpha}\!\!\int_{t_{0}}^{t}\!\!\!d\tau\boldsymbol{g}_{\alpha}(t,\tau)\boldsymbol{u}(\tau,t_{0})=0,
\label{ut}
\end{equation}
and $\boldsymbol{v}(\tau,t)$ is given by\cite{TC2010,TC2012,Yang14}
\begin{equation}
\boldsymbol{v}(\tau,t)=\!\!\int_{t_{0}}^{\tau}\!\!\!d\tau_{1}\!\int_{t_{0}}^{t}\!\!\!d\tau_{2}\sum_{\alpha}\boldsymbol{u}(\tau,\tau_{1})\widetilde{\boldsymbol{g}}_{\alpha}(\tau_{1},\tau_{2})\boldsymbol{u}^{\dagger}(t,\tau_{2}).
\end{equation}
The integral kernel in the above equations are
\begin{subequations}
\begin{align}
& \boldsymbol{g}_{\alpha}(t,\tau)=\int_{-\infty}^{\infty} \frac{d\varepsilon}{2\pi} \boldsymbol{\Gamma}_{\alpha}(\varepsilon) e^{-i \varepsilon (t-\tau)} ,  \\
& \widetilde{\boldsymbol{g}}_{\alpha}(t,\tau)=\int_{-\infty}^{\infty} \frac{d\varepsilon}{2\pi} \boldsymbol{\Gamma}_{\alpha}(\varepsilon) f_{\alpha}(\varepsilon) e^{-i \varepsilon (t-\tau)} ,
\end{align}
\end{subequations}
with
\begin{equation}
\Gamma_{\alpha ij}(\varepsilon)=2\pi \sum_{k} V_{i\alpha k}V_{j\alpha k}^{*}\delta(\varepsilon-\varepsilon_{\alpha k}) ,
\end{equation}
where $\boldsymbol{\Gamma}_{\alpha}(\varepsilon)$ is the spectral density (level broadening) of lead $\alpha$, and
$f_{\alpha}(\varepsilon)=1/[e^{\beta\left(\varepsilon-\mu_{\alpha}\right)}+1]$
is the corresponding Fermi-Dirac distribution function with the chemical potential $\mu_{\alpha}$ and the initial inverse temperature $\beta=1/k_{B}T$.
Because the couplings between dots and leads contain the explicit dependence of the AB phase, we can analyse the AB phase dependence of the reduced density matrix $\rho (t)$ through the master equation (\ref{MasterEq}).

In order to study the molecular states of the DQDs, we change the basis and solve the master equation (\ref{MasterEq}) by diagonalizing $H_{DQD}$.
By labeling the antibonding state (AS) and the bonding state (BS) with the signs $+$ and $-$ respectively, the Hamiltonian of DQDs becomes:
\begin{equation}
H_{DQD}=\sum_{\nu=\pm}\epsilon_{\nu}d_{\nu}^{\dagger}d_{\nu},
\end{equation}
where $\epsilon_{\pm}$ is the corresponding energy level, and $d_{\pm}$ ($d_{\pm}^{\dagger}$) is the corresponding annihilation (creation) operator, which are given by:
\begin{subequations}
\begin{align}
& ~~ \epsilon_{\pm} =  ~ \frac{1}{2}\Big[\left(\epsilon_{11}+\epsilon_{22}\right)\pm\sqrt{\left(\epsilon_{11}-\epsilon_{22}\right)^{2}+4t_{c}^{2}}\Big],
\\
& \begin{pmatrix}d_{+}\\
d_{-}
\end{pmatrix} =  ~\begin{pmatrix} \cos\frac{\theta}{2}  & -\sin\frac{\theta}{2} \\
\sin\frac{\theta}{2} & \cos\frac{\theta}{2}
\end{pmatrix}\begin{pmatrix}d_{1}\\
d_{2}
\end{pmatrix},
\end{align}
\end{subequations}
and $\tan\theta=2t_{c}/(\epsilon_{11}-\epsilon_{22})$.
By denoting the empty state with $\left| 0\right>$, the states AS and BS with $\left| \nu\right>\!\!:\ =\left| \pm\right>$, and the doubly-occupied state by $\left| d\right>$, the reduced density matrix elements at the later time $t$ for an arbitrary initial DQDs state are\cite{me}
\begin{subequations}
\begin{align}
&\rho_{00}(t) = A(t)\Big\{ \rho_{00}(t_{0})+\rho_{dd}(t_{0})det\left[\boldsymbol{J}_{3}(t)\right] \nonumber \\
&~~~~~~~~~~~~~~~~~~~~~~~~~~~~ - \!\!\!\!\!\sum_{\nu,\nu'=: \pm}\!\!\! \rho_{\nu\nu'}(t_{0})
J_{3\nu\nu'} (t) \Big\}, \\
&\rho_{++}(t) = 1\!-\!\rho_{00}(t)\!-\!\rho^{(1)}_{--}(t),~~~\rho_{+-}(t) = \rho^{(1)}_{+-}(t),\\
& \rho_{--}(t) = 1\!-\!\rho_{00}(t)\!-\!\rho^{(1)}_{++}(t),~~~ \rho_{-+}(t) = \rho_{+-}^{*}(t),\\
& \rho_{dd}(t) = 1-\rho_{00}(t)-\rho_{++}(t)-\rho_{--}(t),
\end{align}
\end{subequations}
and the other off-diagonal density matrix elements between the different states are all zero, where
$A(t)=det[\boldsymbol{I}-\boldsymbol{v}(t,t)]$,
$\boldsymbol{J}_{3}(t)=\boldsymbol{u}^\dagger(t,t_{0})(\boldsymbol{I}-\boldsymbol{v}(t,t))^{-1}\boldsymbol{u}(t,t_{0})-\boldsymbol{I}$, and $\boldsymbol{I}$ is the identity matrix,
$\rho(t_{0})$ is initial reduced density matrix, and the single-particle reduced density matrix is given by\cite{TC2010}
\begin{align}
\rho_{\nu\nu'}^{(1)}(t)={\rm Tr}_{S}\big[a_{\nu'}^{\dagger}a_{\nu}\rho(t)\big]=\big[\boldsymbol{u}(t)\rho^{(1)}(t_{0})\boldsymbol{u}^{\dagger}(t) + \boldsymbol{v}(t) \big]_{\nu\nu'}.
\end{align}

\subsection{Quantum transport current}
\label{current derivation}
The transient transport current of electrons flowing from lead $\alpha$ into the DQDs is defined by
\begin{equation}
I_{\alpha}(t)=-e\frac{d}{dt}\mathrm{Tr}_{S\otimes R}\big[ \rho_{tot}(t)N_{\alpha}\big],
\end{equation}
where $N_{\alpha} \equiv\sum_{k}c_{\alpha k}^{\dagger}c_{\alpha k}$.
Using the master equation (\ref{MasterEq}),
the transient current can be expressed in terms of the Green functions
$\boldsymbol{u}(t,t_{0})$ and $\boldsymbol{v}(\tau,t)$:\cite{TC2010,TC2012}
\begin{align}
I_{\alpha}(t) = & -2e\mathrm{Re}\mathrm{Tr}\!\!\int_{t_{0}}^{t}\!\!\!d\tau\Big\{\boldsymbol{g}_{\alpha}(t,\tau)\boldsymbol{u}(\tau,t_{0})\rho^{(1)}(t_{0})\boldsymbol{u}^{\dagger}(t,t_{0})\nonumber \\
 &   +\boldsymbol{g}_{\alpha}(t,\tau)\boldsymbol{v}(\tau,t)-\tilde{\boldsymbol{g}}_{\alpha}(t,\tau)\boldsymbol{u}^{\dagger}(t,\tau)\Big\}.
\label{Ialpha}
\end{align}
This expression of the transient currents can also be derived directly from Keldysh's Green function technique,\cite{TrnCrnt2008} but the dependence of initial conditions, i.e.~the first term in Eq.~(\ref{Ialpha}), was omitted in Ref.~[\onlinecite{TrnCrnt2008}].

According to the analysis used in Ref.~[\onlinecite{DetectParity2011}], one should divide the total transport current into components flowing through AS and BS channels separately.
We can separate the transient current into three parts: the current
component associated with AS or BS channel,
\begin{align}
I_{\alpha\pm}(t)  \equiv & -2e\mathrm{Re}\int_{t_{0}}^{t}d\tau\Big{\{}g_{\alpha\pm\pm}(t,\tau)v_{\pm\pm}(\tau,t)\nonumber \\
 &   +g_{\alpha\pm\pm}(t,\tau)\big[\boldsymbol{u}(\tau,t_{0})\rho^{(1)}(t_{0})\boldsymbol{u}^{\dagger}(t,t_{0})\big]_{\pm\pm}\nonumber \\
 &   -\tilde{g}_{\alpha\pm\pm}(t,\tau)u_{\pm\pm}^{\dagger}(t,\tau)\Big{\}},
 \label{dfnsIaIb}
\end{align}
and the current component associated with both AS and BS channels,
\begin{align}
I_{\alpha+-}(t)  \equiv & -2e\mathrm{Re}\sum_{\nu=\pm}\int_{t_{0}}^{t}d\tau\Big\{g_{\alpha\nu\bar{\nu}}(t,\tau)v_{\bar{\nu}\nu}(\tau,t)\nonumber \\
 &   +g_{\alpha\nu\bar{\nu}}(t,\tau)\big[\boldsymbol{u}(\tau,t_{0})\rho^{(1)}(t_{0})\boldsymbol{u}^{\dagger}(t,t_{0})\big]_{\bar{\nu}\nu}\nonumber \\
 &   -\tilde{g}_{\alpha\nu\bar{\nu}}(t,\tau)u_{\bar{\nu}\nu}^{\dagger}(t,\tau)\Big\},
\end{align}
where $\bar{\nu}$ denotes the opposite sign to $\nu$.
With these definitions, for arbitrary spectral density $\boldsymbol{\Gamma}_{\alpha}(\varepsilon)$,
the transient current through lead $\alpha$ is
\begin{equation}
I_{\alpha}(t)=I_{\alpha+}(t)+I_{\alpha-}(t)+I_{\alpha+-}(t),
\end{equation}
and the transport currents passing from the left to the right leads are
\begin{subequations}
\begin{align}
& I(t)  = \frac{1}{2}\left(I_{L}(t)-I_{R}(t)\right),\\
& I_{\pm}(t) = \frac{1}{2}\left(I_{L\pm}(t)-I_{R\pm}(t)\right),\\
& I_{+-}(t) = \frac{1}{2}\left(I_{L+-}(t)-I_{R+-}(t)\right).
\end{align}
\end{subequations}

After giving the above general formalism, we now solve the problem under the conditions given in Ref.~[\onlinecite{DetectParity2011}], namely,
the energy of each dot $\epsilon_{11}=\epsilon_{22}=\epsilon_{0}$,
the spectral density of lead $\alpha$ $\boldsymbol{\Gamma}_{\alpha}(\varepsilon)=\boldsymbol{\Gamma}_{\alpha}$ (wide band limit) with the level broadenings of the left lead $\Gamma_{L11}=\Gamma_{L22}=\Gamma_{L}$ and the right lead $\Gamma_{R11}=\Gamma_{R22}=\Gamma_{R}$.
The indirect interdot coupling of the left lead $\Gamma_{L12}=a_{L}\Gamma_{L}e^{i\frac{\varphi}{2}}$ and the right lead $\Gamma_{R12}=a_{R}\Gamma_{R}e^{-i\frac{\varphi}{2}}$.
In the molecular basis, the level broadening matrix $\boldsymbol{\Gamma}_{\alpha}$ is given by
\begin{equation}
\boldsymbol{\Gamma}_{L,R}=\Gamma_{L,R}\left(\boldsymbol{I}-\vec{\alpha}_{L,R}\cdot\vec{\boldsymbol{\sigma}}\right),
\end{equation}
where $\vec{\alpha}_{L,R}\!=\!(\alpha_{L,R}^{x},\alpha_{L,R}^{y},\alpha_{L,R}^{z})\!=\!a_{L,R}(0,\mp\sin\frac{\varphi}{2},\cos\frac{\varphi}{2})$ and $\vec{\boldsymbol{\sigma}}$ are the Pauli matrices.
Then the retarded green function has a simple solution,
\begin{eqnarray}
\boldsymbol{u}(t,t_{0})\!\! & = & \!\!
\begin{pmatrix}
u_{++}(t,t_{0}) & u_{+-}(t,t_{0}) \\
u_{-+}(t,t_{0}) & u_{--}(t,t_{0})
\end{pmatrix}\nonumber \\
\!\! & = & \! \exp\Big[\Big(-i\boldsymbol{\epsilon}-\frac{1}{2}\boldsymbol{\Gamma}_{L}-\frac{1}{2}\boldsymbol{\Gamma}_{R} \Big)(t-t_{0})\Big]
\end{eqnarray}
where $\boldsymbol{\epsilon}=\begin{pmatrix} \epsilon_{+}& \!\! 0\\0& \epsilon_{-}\end{pmatrix}$.
For weak indirect interdot coupling, the transport currents associated with the AS or BS channels in the steady-state limit would be the same as those given in Ref.~[\onlinecite{DetectParity2011}],
\begin{equation}
I_{\pm}=\frac{e}{2\pi}\int_{-\infty}^{+\infty}d\varepsilon\big[f_{L}(\varepsilon)-f_{R}(\varepsilon)\big]\Gamma_{L\pm\pm}\Gamma_{R\pm\pm}\left|G_{\pm\pm}^{r}(\varepsilon)\right|^{2}
\label{stdylmtIaIb}
\end{equation}
where $\mathbf{G}^{r}(\varepsilon)=-i\int_{0}^{\infty}e^{i\varepsilon t}\boldsymbol{u}(t)dt,$ which is the retarded Green function in energy domain, and
$\Gamma_{L\pm\pm}\Gamma_{R\pm\pm}\left|G_{\pm\pm}^{r}(\varepsilon)\right|^{2}$ are the effective transmission coefficients of the AS (BS) channels.

\section{The transport current through the AS and BS channels}
\label{s3}

\subsection{The transport current through the AS and BS channels and the justification of the experimental analysis}
\label{s31}

\begin{figure}[t]
\includegraphics[width=8.2cm]{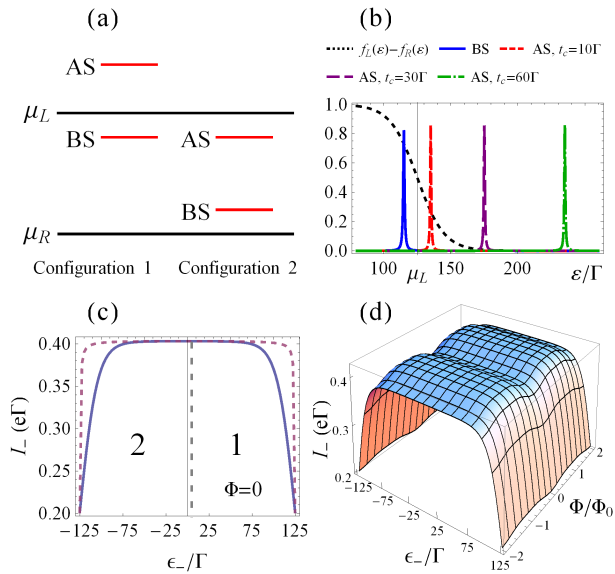}
\caption{
(a) The schematic plot of the AS and BS energy levels in configuration 1 and 2 with the chemical potential of the left and right leads. (b) The difference of the left and right lead particle distributions, $f_{L}(\varepsilon)-f_{R}(\varepsilon)$, and the effective transmission coefficients of the AS and BS channels in configuration 1 for different interdot coupling $t_{c}$ are plotted.
In this case, the BS energy $\epsilon_{-}$ is fixed at $115\Gamma$.
(c) $I_{-}$ as a function of $\epsilon_{-}$ is plotted.
The blue solid line is for temperature $k_{B}T=10\Gamma$, and the purple dashed line is for zero temperature.
The numbers 1, 2 in the plot denote the corresponding energy configurations for $\left| t_{c}\right| =60\Gamma$.
(d) $I_{-}$ is plotted as a function of $\epsilon_{-}$ and $\Phi$.}
\label{fig1}
\end{figure}

In the experiment [\onlinecite{DetectParity2011}], currents under two different energy configurations for AS and BS channels are measured with the fixed bias and indirect interdot weak couplings, as shown in Fig.~\ref{fig1}(a).
Other parameter setting in Ref.~[\onlinecite{DetectParity2011}] are as follow:
the level broadenings of the left lead $\Gamma_{L}=0.3\Gamma$
and the right lead $\Gamma_{R}=0.7\Gamma$ ($\Gamma=\Gamma_{L}+\Gamma_{R}$),
the indirect interdot coupling parameters $a_{L}=-0.1$ for the left lead and $a_{R}=0.15$ for the right lead,
the direct interdot coupling $t_{c}=-60\Gamma$,
the chemical potentials of the left lead $\mu_{L}=125\Gamma$ and the right lead $\mu_{R}=-125\Gamma$,
and the temperature of the reservoirs is set at $k_{B}T=10\Gamma$.
The measured currents are the total currents in each configuration.
As shown by Fig.~\ref{fig1}(a), in configuration 1, only the BS energy state locates within the bias window ($\mu_{L}-\mu_{R}$). In configuration 2, both AS and BS energy states lie in the bias window.
These two energy configurations can be succeeded by tuning gate voltages.
In configuration 1, the current flowing through the BS channel, denoted by $I_{1-}$, is dominant such that the total current is almost given by $I\simeq I_{1-}$, where the current $I_{1+}$ flowing through the AS channel in configuration 1 is negligible.
In configuration 2, the total current $I_{2}=I_{2+}+I_{2-}+I_{2+-}$, where $I_{2+}$, $I_{2-}$ are the currents flowing through the AS and BS channels in configuration 2, respectively, and $I_{2+-}$ is the amount of current flowing through both the AS and BS channels.
The latter is negligible because there is no direct coupling between the AS and BS channels.
Therefore, the total current in configuration 2 is mainly given by $I_{2}\simeq I_{2+}+I_{2-}$.
With the assumption that currents flowing through the BS channel in configuration 1 and 2 are almost the same [\onlinecite{DetectParity2011}], $I_{1-}\simeq I_{2-}$, one can determine the currents flowing through the AS and BS channels, respectively by the total currents measured separately in configuration 1 and 2.
This is the method used in Ref.~[\onlinecite{DetectParity2011}] for analysing the currents flowing through the AS and BS channels.

In the above experimental analysis, we shall check first: (1), whether the current $I_{1+}$ flowing through the AS channel in configuration 1 is really negligible; and (2), what are the conditions that should be satisfied such that the assumption $I_{1-}\approx I_{2-}$ is valid.
According to Eq.~(\ref{stdylmtIaIb}), $I_{1+}$ depends on the overlap of the difference of particle number distributions in the two leads, $f_{L}(\varepsilon)-f_{R}(\varepsilon)$, with the effective transmission coefficient of AS channel, $\Gamma_{L++}\Gamma_{R++}\left|G_{++}^{r}(\varepsilon)\right|^{2}$.
In Fig.~\ref{fig1}(b), the difference $f_{L}(\varepsilon)-f_{R}(\varepsilon)$ is shown by the black dashed line.
We theoretically fix the BS energy, $\epsilon_{-}=\epsilon_{0}-|t_{c}|$, and change the interdot coupling $t_{c}$ to compare the corresponding AS channel contributions to the current.
In experiments, $\epsilon_{-}$ can be manipulated through tuning the energy of DQDs and the interdot coupling simultaneously.
We fix $\epsilon_{-}$ so that the effective transmission coefficient $\Gamma_{L--}\Gamma_{R--}\left|G_{--}^{r}(\varepsilon)\right|^{2}$ of the BS channel is fixed, which is shown by the blue peak in Fig.~\ref{fig1}(b).
Other peaks are the corresponding effective transmission coefficient $\Gamma_{L++}\Gamma_{R++}\left|G_{++}^{r}(\varepsilon)\right|^{2}$ of the AS channel for different $t_{c}$.
As shown by Fig.~\ref{fig1}(b), the larger $t_{c}$ gives the smaller overlap of $f_{L}(\varepsilon)-f_{R}(\varepsilon)$ with $\Gamma_{L++}\Gamma_{R++}\left|G_{++}^{r}(\varepsilon)\right|^{2}$ and hence the smaller current  $I_{1+}$ flowing through the AS channel in configuration 1.
So we conclude that $I_{1+}$ is negligible when $t_{c}$ is properly large enough to make $\Gamma_{L++}\Gamma_{R++}\left|G_{++}^{r}(\varepsilon)\right|^{2}$ lesser overlap with $f_{L}(\varepsilon)-f_{R}(\varepsilon)$.

On the other hand, we plot the current $I_{-}$ flowing through BS channel as a function of the BS energy $\epsilon_{-}$ in Fig.~\ref{fig1}(c).
In our numerical calculation, the parameters are set up according to Ref.~[\onlinecite{DetectParity2011}].
Figure \ref{fig1}(c) shows that the current $I_{-}$ flowing through BS channel becomes maximum when the BS energy $\epsilon_{-}$ is in the middle of the bias window.
Then $I_{-}$ symmetrically and dramatically decays when $\epsilon_{-}$ approaches closely to $\mu_{L}$ or $\mu_{R}$.
Meanwhile, we fix $\left| t_{c}\right| =60\Gamma$ here so that the energy difference of AS and BS is fixed.
Then we can use $\epsilon_{-}$ to determine which energy configuration is examined.
In Fig.~\ref{fig1}(c), the blue solid line gives the current $I_{-}$ as function of $\epsilon_{-}$ for temperature $k_{B}T=10\Gamma$.
It shows that $I_{-}$ is almost a constant within $\left| \epsilon_{-} \right| \lesssim 80\Gamma$.
This indicates that the condition $I_{1-}\simeq I_{2-}$ is well satisfied for $\left| \epsilon_{-} \right| \lesssim 80\Gamma$.
We also show $I_{-}$ at zero temperature in Fig.~\ref{fig1}(c) (the purple dashed line).
In this case, the range for $I_{-}$ being almost a constant is wider.
Also, this flat pattern is maintained for arbitrary magnetic flux $\Phi$ (see Fig.\ref{fig1}(d)).
This ensures that the analysis used in Ref.~[\onlinecite{DetectParity2011}] is valid for all the magnetic flux $\Phi$, and therefore the AB phase dependence of the AS or BS currents can be measured experimentally.

\begin{figure*}[t]
\includegraphics[width=14.12cm]{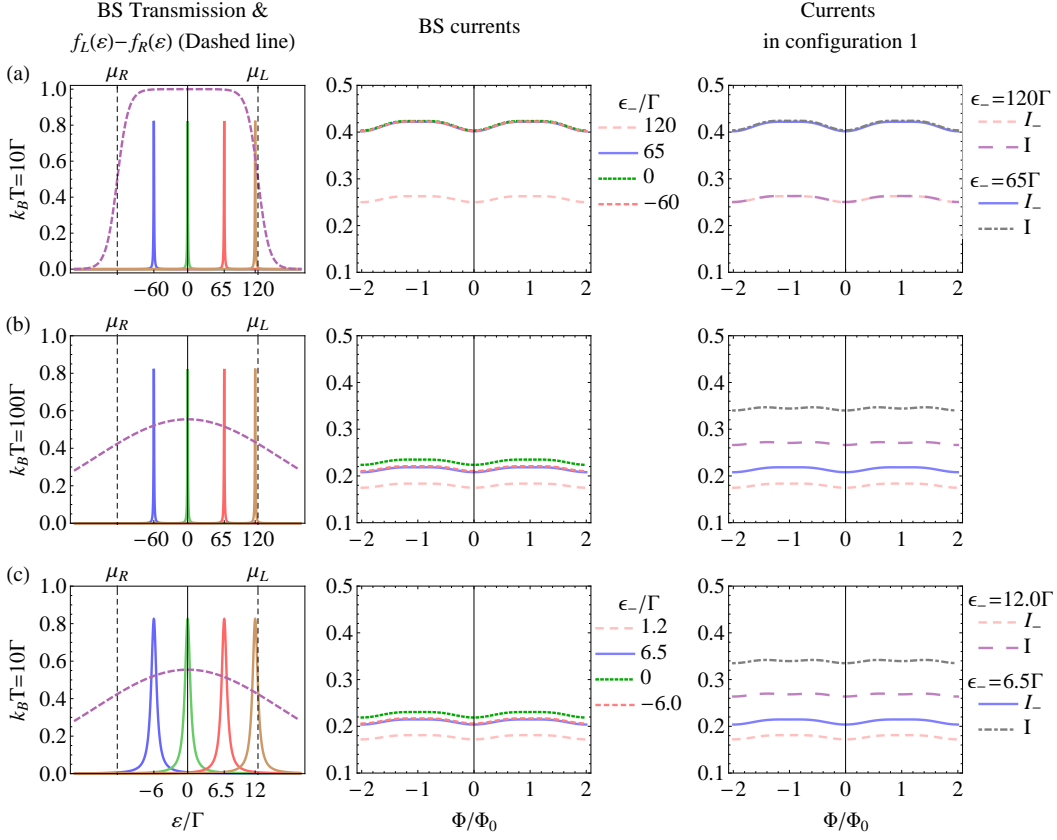}
\caption{The effective transmission coefficient $\Gamma_{L--}\Gamma_{R--}\left|G_{--}^{r}(\varepsilon)\right|^{2}$ of BS and the difference of the left and right leads particle distribution, $f_{L}(\varepsilon)-f_{R}(\varepsilon)$, for different BS energy are plotted in the left side.
The $f_{L}(\varepsilon)-f_{R}(\varepsilon)$ is marked by dashed line.
The corresponding BS currents are plotted in the middle.
The BS currents and the measured currents in configuration 1 ($\epsilon_{-}=65, 120\Gamma$ as examples here) are plotted in the right side.
(a) original settings in Ref.~[\onlinecite{DetectParity2011}]. (b) high temperature case. (c) the stronger coupling case.
}
\label{fig2}
\end{figure*}

Now we should check if this analysis can be applied to other settings.
According to Eq.~(\ref{stdylmtIaIb}), the magnitude of the BS current $I_{-}$ depends on the overlap between the quantities $f_{L}(\varepsilon)-f_{R}(\varepsilon)$ and $\Gamma_{L--}\Gamma_{R--}\left|G_{--}^{r}(\varepsilon)\right|^{2}$.
In Fig.~\ref{fig2}, we choose $\epsilon_{-}=-60, 0, 65, 120\Gamma$ as examples, where $\epsilon_{-}=65, 120\Gamma$ is in configuration 1 and $\epsilon_{-}=-60, 0\Gamma$ is in configuration 2.
Figure~\ref{fig2} gives the overlaps between $f_{L}(\varepsilon)-f_{R}(\varepsilon)$ with $\Gamma_{L--}\Gamma_{R--}\left|G_{--}^{r}(\varepsilon)\right|^{2}$ for these different $\epsilon_{-}$ (see the left column).
The BS currents in configuration 1 and 2 are also shown in Fig.~\ref{fig2} (the central column), and the BS currents and measured currents in configuration 1 are given in the right column in Fig.~\ref{fig2}.
Currents $I_{-}$ for $\epsilon_{-}=-60, 0, 65, 120\Gamma$ are shown by red small dashed line, green dotted line, blue solid line and pink medium dashed line respectively, and currents $I$ for $\epsilon_{-}=65, 120\Gamma$ are shown by black dot-dashed line and purple large dashed line respectively.
Figure \ref{fig2}(a) shows that the analysis works well in the original setting because $f_{L}(\varepsilon)-f_{R}(\varepsilon)=1.0$ when $\left| \epsilon_{-}\right| \lesssim 80\Gamma$.
The overlaps between $f_{L}(\varepsilon)-f_{R}(\varepsilon)$ with $\Gamma_{L--}\Gamma_{R--}\left|G_{--}^{r}(\varepsilon)\right|^{2}$ for different $\epsilon_{-}$ hardly change in this region.
In Fig.~\ref{fig2}(a), $I_{-}$ and $I$ for $\epsilon_{-}=65\Gamma$ are covered by $I_{-}$ for $\epsilon_{-}=-60, 0\Gamma$.
However, when the temperature becomes higher or the couplings of DQDs to leads become stronger, as shown in Fig.~\ref{fig2}(b) and (c) respectively, the overlaps between $f_{L}(\varepsilon)-f_{R}(\varepsilon)$ with $\Gamma_{L--}\Gamma_{R--}\left|G_{--}^{r}(\varepsilon)\right|^{2}$ for different $\epsilon_{-}$ are different.
In Fig.~\ref{fig2}(b), $f_{L}(\varepsilon)-f_{R}(\varepsilon)$ becomes broadened because of the higher temperature, and gives different overlaps with $\Gamma_{L--}\Gamma_{R--}\left|G_{--}^{r}(\varepsilon)\right|^{2}$ and hence the different contribution to $I_{-}$. On the other hand, the broadened $f_{L}(\varepsilon)-f_{R}(\varepsilon)$ due to higher temperature would also overlap with the effective transmission coefficient of AS channel in configuration 1, which makes the measured current $I$ different from $I_{-}$ (see $I_{-}$ and $I$ for $\epsilon_{-}=65, 120\Gamma$ in Fig.~\ref{fig2}(b)).
In Fig.~\ref{fig2}(c), we consider stronger couplings to the leads and still take $\Gamma$ as unit.
The parameters are set as follow: $t_{c}=-6\Gamma$,
 $\mu_{L}=12.5\Gamma$ and $\mu_{R}=-12.5\Gamma$.
In Fig.~\ref{fig2}(c), the stronger couplings to the leads make the level broadenings larger and the transmission coefficients wider.
Then the transmission coefficients of the BS channel result in different $I_{-}$ under different configurations.
In addition, the wider transmission coefficients of AS channel enhance the contribution of the AS current in configuration 1 and thus make $I$ different from $I_{-}$ (see $I_{-}$ and $I$ for $\epsilon_{-}=6.5, 12\Gamma$ in Fig.~\ref{fig2}(c)).
As a result, for higher temperature or stronger coupling to leads, the analysis given in Ref.~[\onlinecite{DetectParity2011}] may become not applicable.

\subsection{Transient transport currents through the AS and BS channels}
\label{s32}
\begin{figure}[t]
\includegraphics[width=8.2cm]{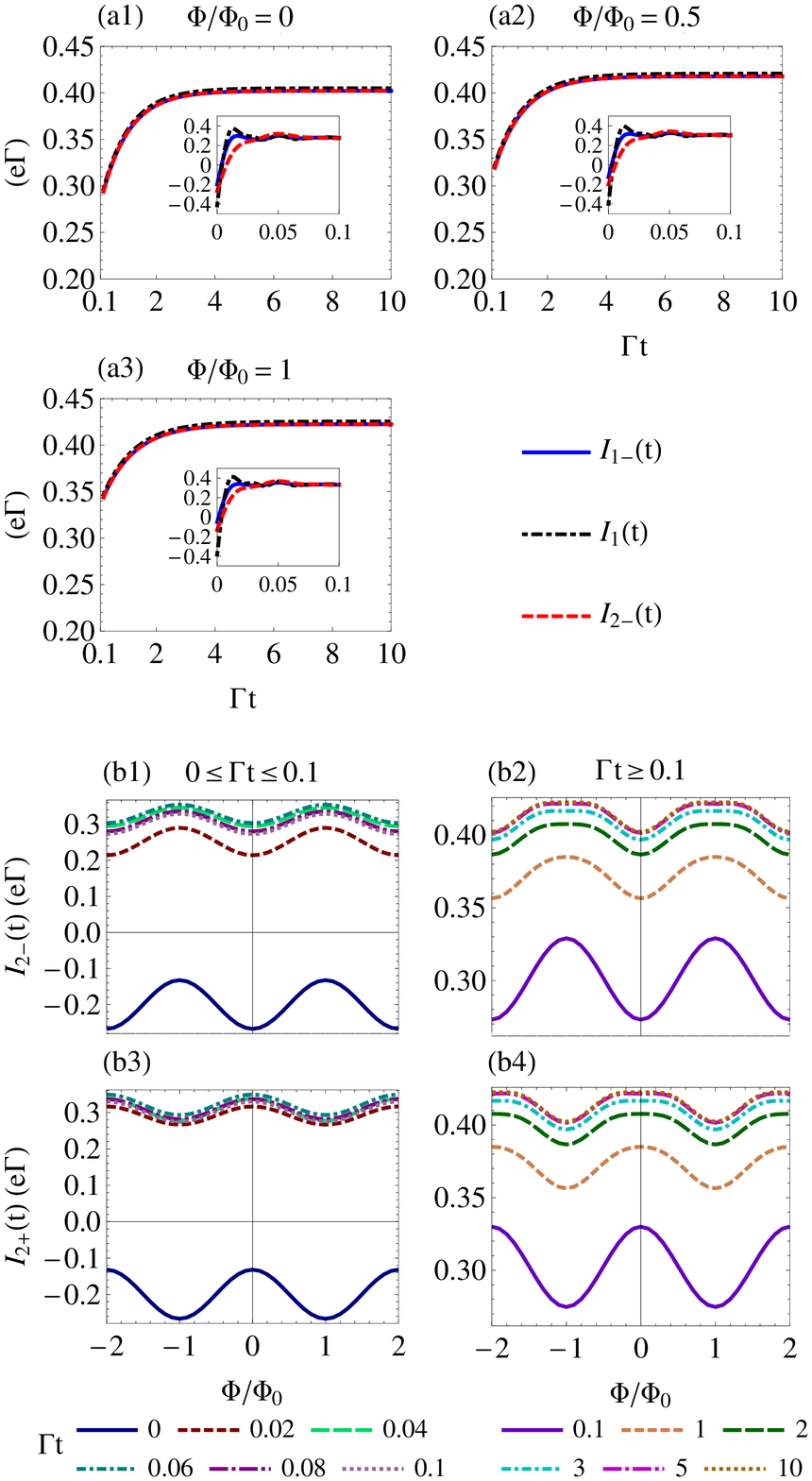}
\caption{
(a) $I_{1-}(t)$, $I_{1}(t)$ and $I_{2-}(t)$ as functions of time for time $\Gamma t\geq 0.1$ under different $\Phi$ are plotted.
The plots for time interval $0\leq \Gamma t \leq 0.1$ are given as insets.
(b) The AB phase dependences of $I_{2-}(t)$ and $I_{2+}(t)$ are plotted.
(b1)-(b2) and (b3)-(b4) respectively for time interval $0\leq \Gamma t \leq 0.1$ and $\Gamma t\geq 0.1$.
}
\label{fig3}
\end{figure}

We have justified the conditions that the currents flowing through BS channel in configuration 1 and 2 are almost the same in the steady-state limit.
It is also interesting to see whether these conditions also exist in the transient regime.
Figures \ref{fig3}(a1)-(a3) show how currents change with time under different magnetic fluxes.
Initially ($0\leq \Gamma t\leq 0.1$), the transient transport current $I_{1-}(t)$ flowing through the BS channel in configuration 1 is not equal to the current $I_{2-}(t)$ in configuration 2.
But they become equal to each other quickly after this short initial time interval.
The difference between $I_{1-}(t)$, $I_{1}(t)$ and $I_{2-}(t)$ in the initial time interval are given as insets in Figs.~\ref{fig3}(a1)-(a3).
Thus, we conclude that the conditions that BS currents in both configurations are almost the same\cite{DetectParity2011} can also be satisfied for the transient regime after a very short time interval from the beginning.

On the other hand, AB phase difference by half a period between the steady AS and BS currents is expected in Ref.~[\onlinecite{DP2004,DP2006}].
Our analysis about AS and BS currents in the steady-state limit also shows this result (see Fig.~\ref{fig4}).
This AB phase difference is thought to be resulted from the parity of AS and BS channels, which is a property of the device geometry.
Because it only depends on the device geometry, the dynamics of the AS and BS channels should not influence this phase difference.
In Figs.~\ref{fig3}(b1)-(b4), $I_{2-}(t)$ ($I_{2+}(t)$) for time interval $0\leq \Gamma t\leq 0.1$ and $\Gamma t \geq 0.1$ are given.
It shows that the AB phase dependence of the AS and BS currents are fixed for all time.
The dynamics of the AS and BS channels does not influence their AB phase difference, as we expected.

\section{Reduced density matrix elements and the corresponding transport currents}
\label{s33}

\begin{figure}[t]
\includegraphics[width=8.2cm]{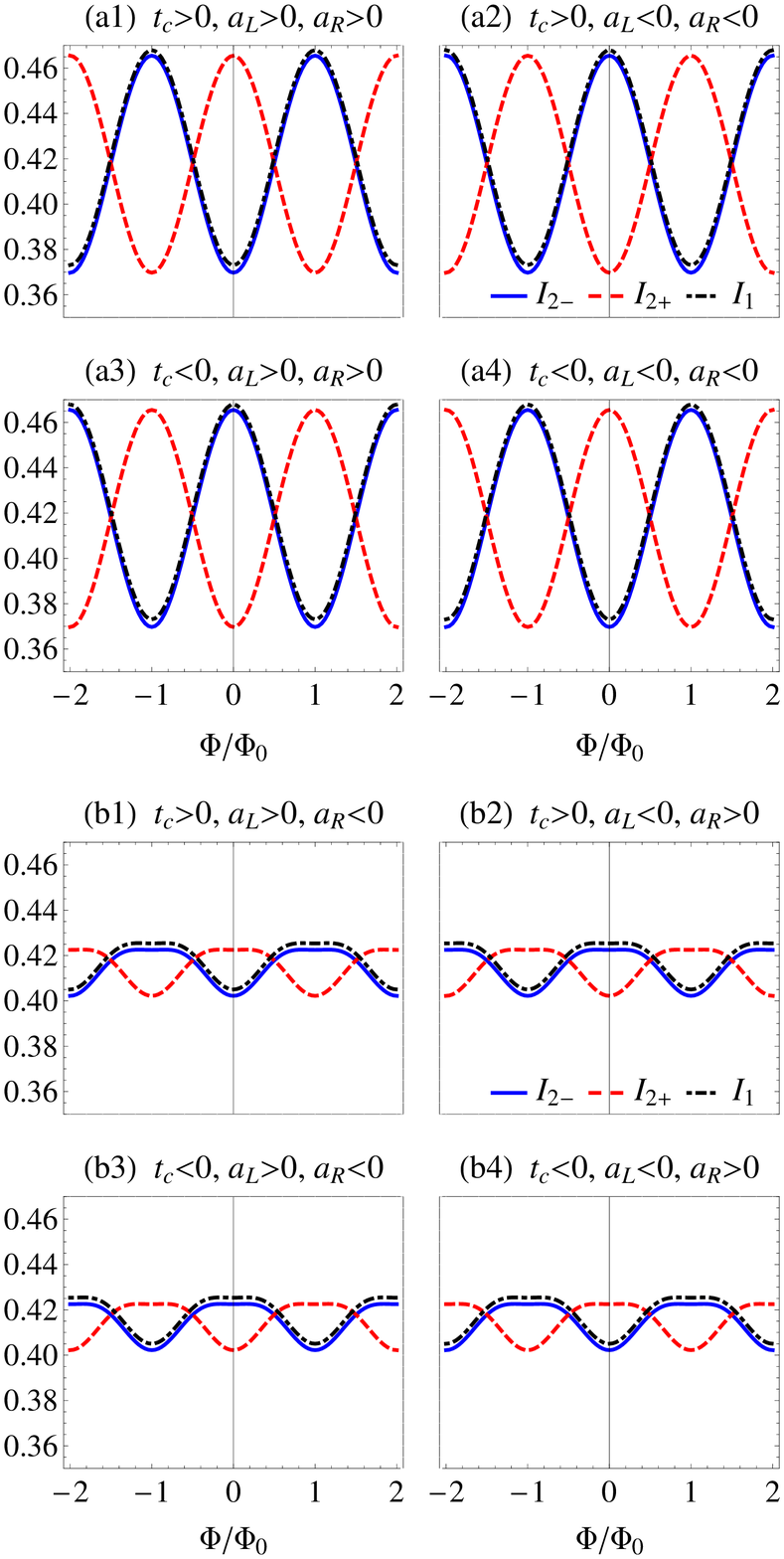}
\caption{
The AB phase dependences of $I_{2-}$, $I_{2+}$ and $I_{1}$ are plotted in unit $e\Gamma$ for different sign configuration of $t_{c}$, $\alpha_{L}$ and $\alpha_{R}$.
The other parameters are taking according to Ref.~[\onlinecite{DetectParity2011}].}
\label{fig4}
\end{figure}

\begin{figure}[t]
\includegraphics[width=8.2cm]{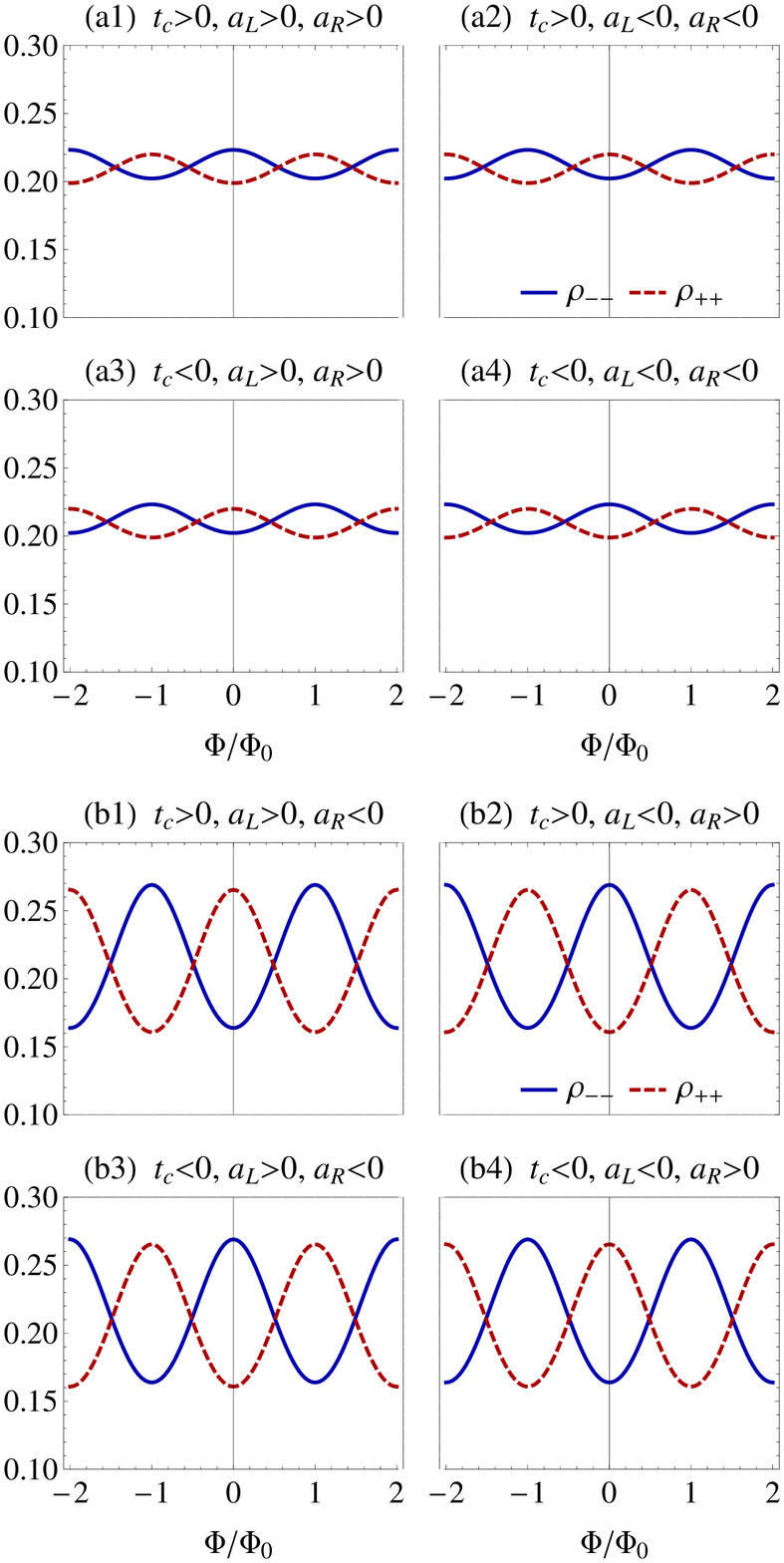}
\caption{
The AB phase dependences of $\rho_{++}$ and $\rho_{--}$ for different signs of $t_{c}$, $\alpha_{L}$ and $\alpha_{R}$.
The top (bottom) side is for $\rho_{++}$ ($\rho_{--}$), and the left (right) side is for $\alpha_{L}\alpha_{L}>0$ $(<0)$.
}
\label{fig5}
\end{figure}

In fact, the device geometry determines not only the parity of AS and BS channels but also the corresponding wave functions.
In this section, we discuss the AB phase dependence between the probabilities of single-electron occupying AS and BS states under different device geometries, which is useful for quantum state tomography.
The AS and BS reduced density matrix elements $\rho_{++}$ and $\rho_{--}$ in the steady-state limit are shown in Fig.~\ref{fig5}.
The different device geometries are controlled by different signs of the direct interdot coupling $t_{c}$ and indirect coupling parameters $a_{L},a_{R}$.
As we see in Figs.~\ref{fig4} and \ref{fig5}, the patterns of AB phase dependences of $\rho_{++}$, $\rho_{--}$ and the corresponding currents for $a_{L}a_{R}<0$ and $a_{L}a_{R}>0$ are different.
Note that when the signs of $a_{L}$ and $a_{R}$ are the same, the level broadenings of the left or right lead coupled to the AS or BS states, which are given by $\Gamma_{\alpha\pm\pm}\!=\Gamma_{\alpha}(1\mp a_{\alpha}\cos\frac{\varphi}{2}), \alpha\!\!=\!\!L$ or $R$, enhance or shrink simultaneously for different magnetic flux.
Consequently, the AB oscillation amplitude of the transport current, which relies $\Gamma_{L\pm\pm}\Gamma_{R\pm\pm}$ (see Eq.~(\ref{stdylmtIaIb})), is sensitive to the flux.
However, when the signs of $a_{L}$ and $a_{R}$ are different, the AS or BS level broadenings is enhanced only for one of the two leads.
Therefore, the AS or BS current becomes less sensitive to the flux.
The enhancement of level broadening only for one of two leads forces electrons to locaize at AS or BS state so that $\rho_{++}$ and $\rho_{--}$ becomes more sensitive to the flux.
This leads to the smaller amplitude of the AB oscillation in the corresponding currents and the larger amplitude in $\rho_{++}$ and $\rho_{--}$ (see Fig.~\ref{fig5}).
On the other hand, the corresponding changes of the AB phase dependences under different geometries are the same.
When the sign of $t_{c}$ or the parameters $a_{L}$ and $a_{R}$ change, the AB phase dependence of $\rho_{++}$ and $\rho_{--}$ will also change accordingly.
From the AB phase dependences of $\rho_{++}$, $\rho_{--}$ and the corresponding currents, we can know the signs of $a_{L}$ and $a_{R}$.

\begin{figure}[t]
\includegraphics[width=8.2cm]{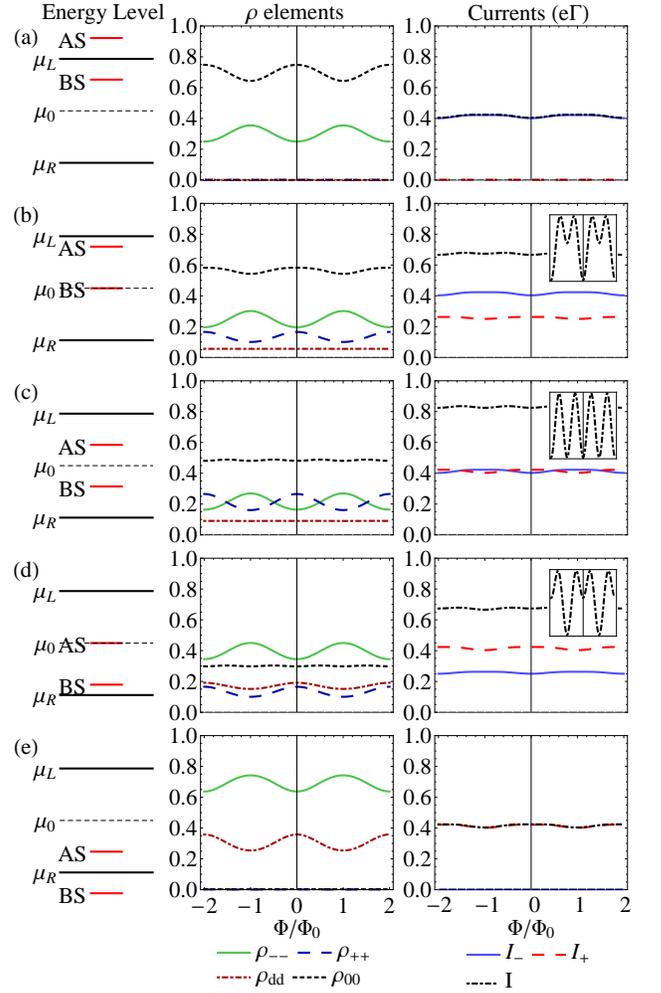}
\caption{
The corresponding AB phase dependences of the reduced density matrix elements and the currents $I_{-}$, $I_{+}$ and $I$ for different energy configuration.
(a)$\epsilon_{-}=65\Gamma$.
(b)$\epsilon_{-}=0\Gamma$.
(c)$\epsilon_{-}=-60\Gamma$.
(d)$\epsilon_{-}=-120\Gamma$.
(e)$\epsilon_{-}=-185\Gamma$. }
\label{fig6}
\end{figure}

On the other hand, the magnitudes of reduced density matrix elements and the AS and BS currents depend sensitively on the AS and BS energy level space and the bias.
We compare such dependence in Fig.~\ref{fig6}.
The relative positions of the AS and BS energy levels and the bias can be roughly separated into five regions:
(a), only BS is inside the bias window;
(b), BS and AS lie between $\mu_{L}$ and $\mu_{0}$, where $\mu_{0}$ is the middle point of the bias window;
(c), the BS and AS levels are between $\mu_{L}$ and $\mu_{0}$ and $\mu_{R}$ and $\mu_{0}$, respectively;
(d), BS and AS lie between $\mu_{R}$ and $\mu_{0}$;
and (e), only AS is inside the bias window, as shown in the plots on the left-most column of Figs.~\ref{fig6}(a)-(e) respectively.
In Fig.~\ref{fig6}(a), when the AS level is outside the bias window, the total current is just the BS current.
The dominant reduced density matrix element is the vacuum state $\rho_{00}$.
The second dominant one is $\rho_{--}$ because the chemical potential of the left lead is larger than BS energy level, which makes electrons tend to hop from the left lead into the BS state.
In Fig.~\ref{fig6}(b), the AS current also contributes to the total current, but the BS current contributes larger because it is closer to $\mu_{0}$.
Therefore, the total current and the BS current would be in phase but their magnitudes are different.
In this case, electrons start to hop into the AS but the dominant elements are still given by the $\rho_{00}$ and $\rho_{--}$.
In Fig.~\ref{fig6}(c), AS and BS currents give similar contributions to the total current such that the period of the AB phase dependence of the total current is approximately half of those of AS and BS currents.
Meanwhile, the largest reduced density matrix element is still $\rho_{00}$, but $\rho_{--}$ is comparable to $\rho_{++}$.
Although $\rho_{00}$ is the mostly occupied state in these three cases, its magnitude becomes lesser and lesser when the energy levels of AS and BS are lowered more and more.
This is because the distance between BS level and $\mu_{R}$ becomes smaller such that electrons become stable in the BS level.
As a result, in the region (d) as shown by Fig.~\ref{fig6}(d), the dominant matrix element becomes $\rho_{--}$ and the AS current also becomes the largest one.
When the BS level is lower than $\mu_{R}$, electrons tend to fill the BS level, which makes $\rho_{--}$ more dominate, see Fig.~\ref{fig6}(e).
In this case, only the AS current contributes to the total current.
The traveling electrons through the AS channel and the stable electrons located in BS level make $\rho_{dd}$ become the second dominant matrix element.
In brief, the magnitudes of AS or BS currents depend on how close the corresponding energy level and $\mu_{0}$ are.
The magnitudes of reduced density matrix elements depend on how close the energy level of BS and $\mu_{R}$ are.
From these analysis, we can determine from the AB phase dependence of the transport current which state the DQDs system stays.
which should provide useful information for the reconstruction of quantum states through the measurement of the transport current.

\begin{figure}[t]
\includegraphics[width=5.5cm]{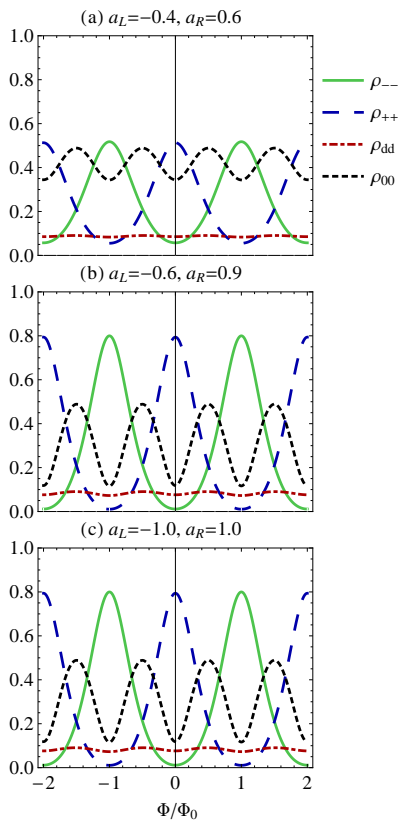}
\caption{
The AB phase dependences of the reduced density matrix elements when $\epsilon_{-}=-60\Gamma$.
The different magnitudes of $a_{L,R}$ are taken here for the larger amplitudes of the AB phase dependences.
(a) $a_{L}=-0.4$, $a_{R}=0.6$
(b) $a_{L}=-0.6$, $a_{R}=0.9$
(c) $a_{L}=-1.0$, $a_{R}=1.0$.
The other parameters are taken according to Ref.~[\onlinecite{DetectParity2011}].}
\label{fig7}
\end{figure}

If the AB phase dependence of the total current is half period of that of the AS or BS current, the AS and BS energy levels are equally close to $\mu_{0}$.
Thus, the traveling electrons passing through these channels have similar contribution to the corresponding reduced density matrix elements.
The magnitudes of $\rho_{++}$ and $\rho_{--}$ are close to each other in this condition (see Fig.~\ref{fig6}(c)).
The larger $\alpha_{L,R}$ makes the electrons that flow from the left lead to the right lead more sensitive on the magnetic flux, and thus gives the larger amplitudes to the reduced density matrix elements $\rho_{++}$ and $\rho_{--}$ and the corresponding currents, as shown in Fig.~\ref{fig7}, where the strong interference between the $\rho_{++}$ and $\rho_{--}$ is also shown.
The DQDs system would have large probability to be in bonding state or antibonding state at certain $\Phi$.
For example, the DQDs system mainly locate in antibonding state (bonding state) at $\Phi=0$ ($\Phi/\Phi_{0}=1$).
Therefore, if the strong indirect interdot coupling can be practically succeeded, it could provide an efficient way to manipulate coherence between the bonding and antibonding states through the magnetic flux.
A similar scheme for DQDs without direct interdot coupling has been proposed in Ref.~[\onlinecite{MnltDSwF2011,MnltDSwF2012}] for manipulating the relative phase between the two QDs.
However, the direct interdot coupling can totally manipulate the energy levels of bonding and antibonding states so that we can distinguish the contributions of the corresponding channels in the transport current.

\section{Summary}
\label{s4}
In conclusion, using the quantum transport theory based on the master equation\cite{TC2010,TC2012,CGFSF2008}, we have justified the method used in Ref.~[\onlinecite{DetectParity2011}] for analysing the AS and BS currents.
We show that when the energy level of BS is within a quite large range near the middle of the bias window, the BS current in configuration 1 well approximates the BS current in configuration 2.
However, in conditions that the temperature is high or the couplings to the environments is strong, this analysis may not be valid.
This is because the bias window is broaden in high temperature and the effective transmissions of the channels become wider for the strong couplings to the leads.
We also extend this current analysis through AS and/or BS channels to the transient regime.
We find that the analysis in Ref.~[\onlinecite{DetectParity2011}] is still valid in the transient regime except for a short time interval from the very beginning.
We also show that the AB phase dependence of these current components is mainly determined by the device geometry, and it is independent of the dynamics of the AS and BS channels.
Furthermore, we also examine the relation between the AB phase dependence of single-electron probabilities of occupying AS and BS and the corresponding currents.
We find that the AB phase dependence varies in the same way under different device geometries.
We also explore the AB phase dependence of the AS and BS reduced density matrix elements and the corresponding currents under different AS and BS energies.
This provides not only the useful information for the reconstruction of quantum states through the measurement of transport current, an approach known as quantum state tomography, but also a practice way to manipulate coherence between the AS and BS states with the magnetic flux, as a key for quantum information processing in quantum dot systems.

\begin{acknowledgements}
We thank to A. Aharony and O. Entin-Wohlman for the fruitful discussions.
This work is supported by the Ministry of Science and Technology, Taiwan, ROC, under Contract No.
NSC102-2112-M-006-016-MY3, by the Headquarters of University
Advancement at the National Cheng Kung University,
which is sponsored by the Ministry of Education, Taiwan,
ROC, and from the National Center for Theoretical Science
, Taiwan and the High Performance Computing Facility in
the National Cheng Kung University.
\end{acknowledgements}

\end{document}